\documentclass[preprint,12pt, a4paper]{elsarticle}

\usepackage{listings}
\usepackage{xcolor}

\usepackage{amssymb}
\usepackage{hyperref}
\usepackage{xurl}
\usepackage{parskip}
\journal{SoftwareX}

\begin{document}
\renewcommand{\labelenumii}{\arabic{enumi}.\arabic{enumii}}

\title{GPS-2-GTFS: A Python package to process  and transform raw GPS data of public transit to GTFS format}

\hypersetup{
    pdftitle={GPS-2-GTFS: A Python package to process and transform raw GPS data of public transit to GTFS format},
    pdfauthor={Shiveswarran Ratneswaran, Uthayasanker Thayasivam, Sivakumar Thillaiambalam},
    pdfsubject={Public Transit, GTFS, GPS Data Processing},
    pdfkeywords={GPS data, GTFS, Public Transit, Trip extraction, Intelligent Transportation Systems}
    }


\author[label1]{Shiveswarran Ratneswaran\corref{cor1}}
\cortext[cor1]{Corresponding author}
\ead{shiveswarran.22@cse.mrt.ac.lk}
\author[label1]{Uthayasanker Thayasivam}
\author[label2]{Sivakumar Thillaiambalam}

\address[label1]{Department of Computer Science and Engineering, University of Moratuwa, Katubedda, Moratuwa, 10400, Sri Lanka}
\address[label2]{Department of Transport Management and Logistics Engineering, University of Moratuwa, Katubedda, 10400, Sri Lanka}

\begin{frontmatter}
\begin{abstract}
The `gps2gtfs' package addresses a critical need for converting raw Global Positioning System (GPS) trajectory data from public transit vehicles into the widely used GTFS (General Transit Feed Specification) format. This transformation enables various software applications to efficiently utilize real-time transit data for purposes such as tracking, scheduling, and arrival time prediction. Developed in Python, `gps2gtfs' employs techniques like geo-buffer mapping, parallel processing, and data filtering to manage challenges associated with raw GPS data, including high volume, discontinuities, and localization errors. This open-source package, available on GitHub and PyPI, enhances the development of intelligent transportation solutions and fosters improved public transit systems globally.
\end{abstract}


\end{frontmatter}


\section*{Metadata}
\label{}
\begin{table}[!htbp]
\begin{tabular}{|l|p{6cm}|p{5.5cm}|}
\hline
\textbf{Nr.} & \textbf{Code metadata description} & \textbf{Please fill in this column} \\
\hline
C1 & Current code version & gps2gtfs 0.1.0 \\
\hline
C2 & Permanent link to code/repository used for this code version & \url{https://github.com/aaivu/gps2gtfs} \\
\hline
C3  & Permanent link to Reproducible Capsule & -\\
\hline
C4 & Legal Code License   & MIT License \\
\hline
C5 & Code versioning system used & git\\
\hline
C6 & Software code languages, tools, and services used & 
Python, geopandas , folium\\
\hline
C7 & Compilation requirements, operating environments \& dependencies & \url{https://github.com/aaivu/gps2gtfs/blob/master/requirements.txt}\\
\hline
C8 & If available Link to developer documentation/manual & \url{https://github.com/aaivu/gps2gtfs/blob/master/PACKAGE_DESCRIPTION.md} \\
\hline
C9 & Support email for questions & rtuthaya@cse.mrt.ac.lk\\
\hline
\end{tabular}
\caption{Code metadata}
\label{codeMetadata} 
\end{table}

\begin{table}[!htbp]
\begin{tabular}{|l|p{6cm}|p{5.5cm}|}
\hline
\textbf{Nr.} & \textbf{(Executable) software metadata description} & \textbf{Please fill in this column} \\
\hline
S1 & Current software version & 0.1.0 \\
\hline
S2 & Permanent link to executables of this version  & For example: \url{https://pypi.org/project/gps2gtfs/} \\
\hline
S3  & Permanent link to Reproducible Capsule & -  \\
\hline
S4 & Legal Software License & MIT License \\
\hline
S5 & Computing platforms/Operating Systems & OS Independent\\
\hline
S6 & Installation requirements \& dependencies & \\
\hline
S7 & If available, link to user manual - if formally published, include a reference to the publication in the reference list & \url{https://pypi.org/project/gps2gtfs/ } \\
\hline
S8 & Support email for questions & rtuthaya@cse.mrt.ac.lk\\
\hline
\end{tabular}
\caption{Software metadata}
\label{executabelMetadata} 
\end{table}

\newpage

\section{Motivation and significance}

The public transit monitoring and management system is a crucial and fundamental component of strategic planning by transport authorities. Indeed, it regulates and creates a convenient and efficient mode of choice for people's mobility. Nevertheless, there should be a real-time data collection system (e.g., sensors) and a framework that processes the raw data into a usable format for downstream applications (e.g., tracking and monitoring \cite{ratneswaran2023extracting}, scheduling \cite{bie2021optimization}, arrival time prediction \cite{ratneswaran2023improved}, behavioral study, etc.). Certainly, the widespread use of GPS (Global Positioning System) sensors in public transit vehicles generates geospatial temporal data abundantly \cite{fan2015dynamic}. However, there is no such library or open package specifically tailored for processing transit GPS trajectory data into the most commonly utilized form of GTFS (General Transit Speed Feed Specification) data that can be consumed by a wide variety of software applications. Today, thousands of public transportation providers worldwide use the GTFS data format to convey information to passengers \cite{antrim2013many}. In this paper, we present the `gps2gtfs' package, which is designed explicitly to transform transit GPS data into GTFS real-time format under many potentially challenging circumstances.
\par
The GTFS data format is widely used and accepted as a standard for transit data sharing. GTFS is comprised of two primary components: GTFS Schedule and GTFS Real-time. The GTFS Schedule comprises data on routes, schedules, fares, and spatial transit features, presented in easily readable text files. Meanwhile, GTFS Real-time encompasses information on real-time updates on vehicle locations, arrival predictions, and service alerts. GTFS-RT data is crucial to generate as it needs to be updated in real-time and relies more on transit planning. Starting in 2023, the Federal Transit Administration of the United States has mandated that transit agencies in the country include accurate GTFS data when submitting their annual National Transit Database report. This will induce other developed and developing nations to prioritize accurate and standardized transit data reporting to improve their public transportation systems. Moreover, GTFS applies to any form of public transit, including buses, trains, subways, metros, trams, ferries, and LRT (light rail transit). 
\par
Meanwhile, several transit authorities require GTFS transit data, primarily to integrate with Google Maps \cite{antrim2013many}. There are, however, many transit and multimodal software applications that can use GTFS data. These include multimodal trip planning, schedule or timetable generation, mobile and web apps for both passengers and authorities, real-time big data visualization of urban transit maps, accessibility, planning analysis tools, real-time information, and interactive voice response (IVR) for alerts and delays \cite{antrim2013many}. Moreover, there is always a need for a framework processing GPS data for advanced traveler information systems (ATIS), for example, the application of tracking and predicting arrival times of buses to all stops along the way \cite{ratneswaran2023improved}.
\par
Nevertheless, the complex nature of mapping raw GPS data to the GTFS format is a significant hurdle, as there is no native library that facilitates this conversion effortlessly. GTFS data is widely utilized by transit agencies and authorities for various applications, making the accuracy and reliability of this transformation crucial. Typical data problems include GPS data's high volume and velocity, necessitating scalable, multi-threaded processing by chunking the data.  Moreover, there are various challenges in this transformation process. First, GPS data are a stream of transit location moving information. When processing, discontinuities or non-uniformities may arise due to inadequate network coverage, particularly in areas with steep terrain. Second, they can be mapped to specific locations on the digital map, although with limited precision. Third, the decreased frequency (i.e., higher time interval in between successive GPS records) of GPS data significantly impacts the accuracy of matching with bus stops buffer zone (localization to points of interest). Our approach ensures efficient handling and processing, while buffer filters are employed to manage and correct errors in localization, enhancing the overall processed information quality.
\par
In previous years, few related software packages have been developed to process the GPS data, which are collected on a massive scale. The `OD Means' \cite{HEREDIA2024101732} is an R-based package to process GPS data collected especially from taxis to understand the mobility patterns of people. Moreover, a software package named `T-Ridership' \cite{IMANI2023101350} was developed for public transportation; nevertheless, its motive is to find optimal bus stations in scheduling. Hence, no software is available to process and transform the public transit GPS data into usable format.

\par
Essentially, our `gps2gtfs' Python library offers an efficient method for processing raw GPS trajectory data and turning it into the GTFS data format. This package utilizes the capabilities of Python Pandas DataFrame and GeoDataFrame, mapping to localization with geo-buffer technique and along with parallelization, to provide effective techniques for extracting crucial trip information from the raw GPS data. The information includes sequences of trip trajectory, transit stop details, arrival and departure times at stops, dwell time at stops (i.e., waiting time to drop off and pick up passengers), trip travel duration, running times between stops (i.e., segment run times), and smooth integration into the GTFS data format. The `gps2gtfs' tool delivers both static trip (schedule) data and dynamic real-time travel data (i.e., trip updates) in varying traffic conditions. 
\par
Additionally, the extensible nature of the `gps2gtfs' library allows for plug-and-play interfaces that can be tailored to specific needs, making it a versatile tool for different downstream tasks. The preprocessing capabilities of this package significantly reduce the burden for future works by mediating the raw data to a cleaned and structured GTFS format. This process not only simplifies data handling but also facilitates error adjustments such as map matching and localization to points of interest. The ability to preprocess and clean data effectively means that subsequent analyses and applications can be more accurate and efficient. The rest of this paper explains the software description including methodological framework and software architecture.

\section{Software description}

This section elaborates the software description including the methodology in developing the `gps2gtfs' software package and the software architecture and its functionalities.

\subsection{Methodology}

The high-level flowchart of the methodology of the processes behind this software is shown in the figure \ref{fig:modelarchi} below. 
\begin{figure}[htbp]
    \centering
    \includegraphics[width = 0.8\textwidth]{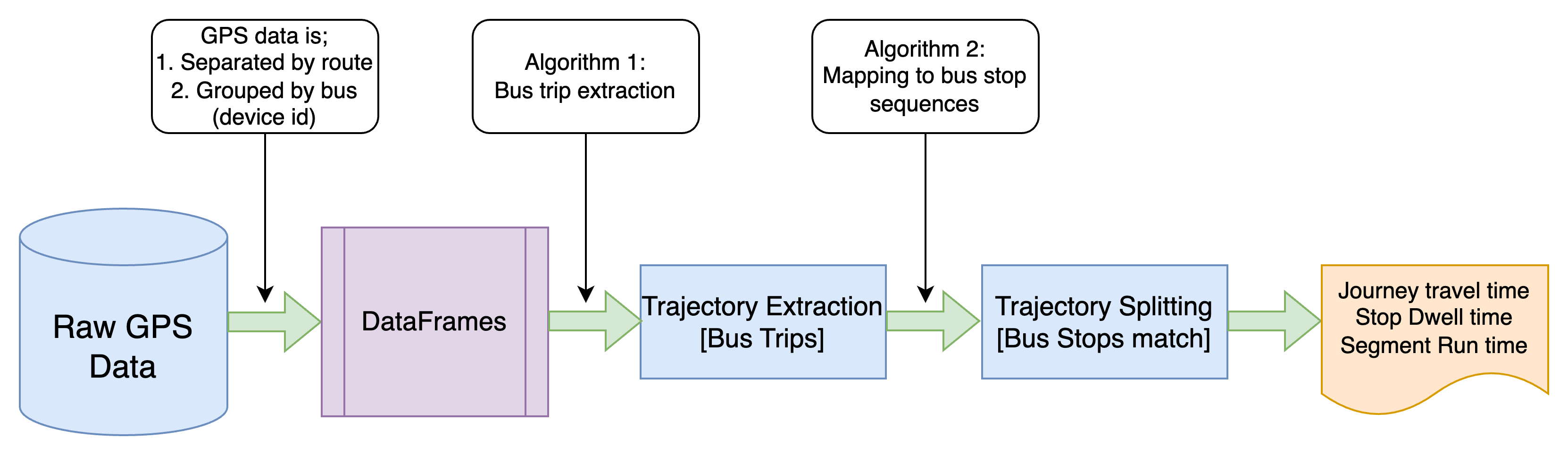}
    \caption{High-level flowchart of the methodology of the `gps2gtfs' package which delivers GTFS information of trip travel times, stop dwell times, and segment run times of each transit trip}
    \label{fig:modelarchi}
\end{figure}

Initially, the input data is gathered using the AVL (Automatic Vehicle Location) system. Such that, every vehicle in the public transit system is equipped with GPS sensors that are connected to GSM modems, and each has a unique device identification code. These sensors generate signals at a predetermined sample rate (frequency) and transmit the data to servers situated in a remote office. The sampling rate heavily influences the quality of this transformation process. If the data emission is less frequent, the dwell time information will not be as accurate. In contrast, high-frequency data collection requires more storage space and processing speed. Our package considered these in their functions. Next, every GPS observation has spatio-temporal characteristics like the latitude and longitude values of the location with timestamps. However,  there may be sensor or precision errors, discontinuities, or nonuniformities in the data signals. Hence, we have developed a module to preprocess the crude raw GPS data before sending it to the main modules. Hereafter, GPS records were grouped by device identifier (each bus) and generated data frames for the following transformation process. There are two main components or algorithms in the transformation process. First, trajectory extraction for each trip, then trajectory splitting to match the bus stop information. These components contribute in estimating the running time and dwell time information, respectively, required for the GTFS format. The following subsections describe the two algorithms.

\subsubsection{Transit trip extraction}

The next major component in the software is the `trajectory extraction'. The pseudocode of the transit trip trajectory extraction algorithm is shown in the figure \ref{fig:algo} below.

\begin{figure}[htbp]
    \centering
    \includegraphics[width = 0.7\textwidth]{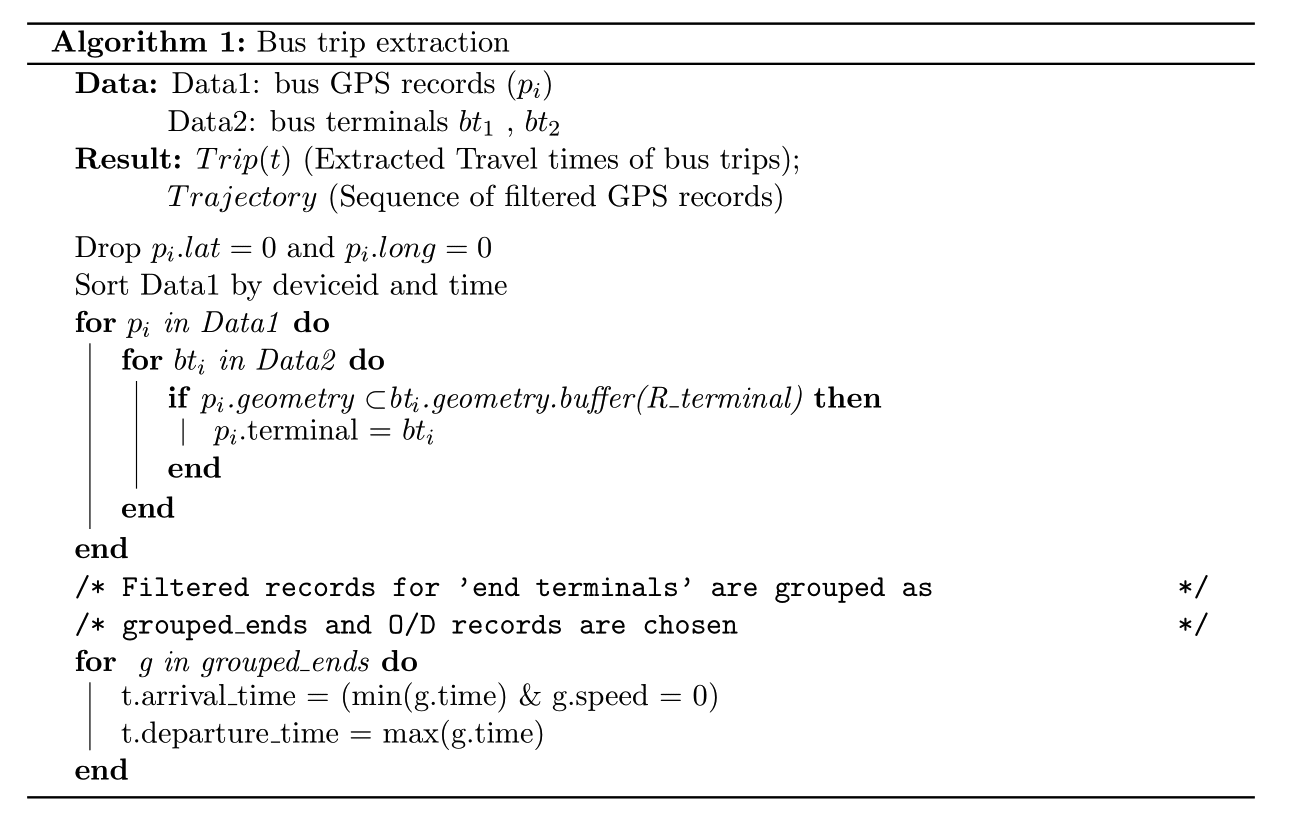}
    \caption{The pseudo-code of the bus trip extraction algorithm which outputs the trip start time and end time along with trip GPS trajectory sequences using the raw GPS data and location metadata of terminals as input }
    \label{fig:algo}
\end{figure}

The input data for this algorithm is the GPS traces of buses and the bus terminal location data. The GPS data that we are receiving from the server is a bunch of many buses in a metropolitan area. They are sorted by routes at first, then grouped by device identifier, which indicates a unique bus, and the end data frame is sorted by time to ensure continuous trip data. GPS sensors generate signals continually when the engine is on and terminate when the engine is off. In reality, drivers wait in the terminal for their next return journey without turning off the engine. The system cannot separate one trip from the succession of GPS recordings. Hence, this algorithm facilitates separate trajectories. The collected GPS records of buses are split into trajectories that depict bus runs from the start terminal to the end terminal. Coordinates for both end terminals should define every valid trip. Positioning inaccuracies are inherent in GPS data. Therefore, identifying records with similar origin or destination terminal locations is impossible. Bus terminals typically indicate an area rather than a point. To identify matched coordinates in space, a buffer area of circular around a point is employed, with the radius of the buffer area (\textit{R$_{terminal}$}). This parameter can be set at the beginning of running this script. The start time (\textit{t$_{start}$}) of a trip is the last observed time point within the buffer region of the origin bus station, whereas the end time (\textit{t$_{end}$}) is the first observed time point of zero speed \textit{(s=0)}. Finally, the trajectories of every bus trip can be extracted. 

\subsubsection{Transit stop matching}

The succeeding component of this software pipeline is `transit stop matching', which is more complex and affected by the data quality parameters such as GPS data emitting frequency and continuities of the data. Bus stop sequences (i.e., corresponding timestamps (data points) for the bus arrival to one-stop, departure from that stop, arrival to the next stop, and continuing until it reaches the terminal) are matched from the continuous emitting GPS data trajectory points using this module `transit stop matching'. When tracking the continuous GPS data stream, when the speed becomes zero near a bus stop, it is taken as the arrival, and the next immediate non-zero speed value record can be taken as the departure from the stop. Here, a buffer radius is (\textit{R$_{stop}$}) defined (e.g., 50m) around every bus stop, and the GPS points within the buffer area are matched to the corresponding stop. However, due to the GPS data quality, there will be some limitations in this process. 

\begin{figure}[htbp]
    \centering
    \includegraphics[width = 0.3\textwidth]{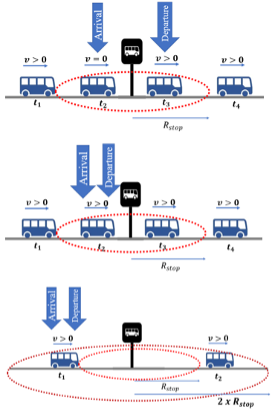}
    \caption{Three potential various scenarios when capturing GPS records within
the buffer area centered around localized bus stops to find out timestamps corresponding to arrival and departure of bus}
    \label{fig:dwell}
\end{figure}

Mainly, as shown in the figure \ref{fig:dwell}, there are three scenarios to take into consideration when matching the GPS record to the arrival and departure at the stop. First, the usual case is that the arrival time is the first \textit{Speed = 0} timestamp, and the successive record with \textit{\(Speed > 0\)}, filtered inside the buffer, is the departure. Second, the case where there is no \textit{Speed = 0} records inside the buffer. This indicates the bus might not have stopped (skipped) or is missing GPS data (e.g., poor network coverage). However, we require the arrival estimate to the stop; hence, the timestamp corresponds to the very close proximity to the stop that is taken as the arrival and departure times, and consequently, dwell time is estimated as 0. Third, there are no filtered records in the current buffer radius. Here, the buffer radius is enlarged (e.g., 100 m), and the closest point is taken for the estimation of arrival time.

\subsection{Software architecture}

\begin{figure}[htbp]
    \centering
    \includegraphics[width = 0.9\textwidth]{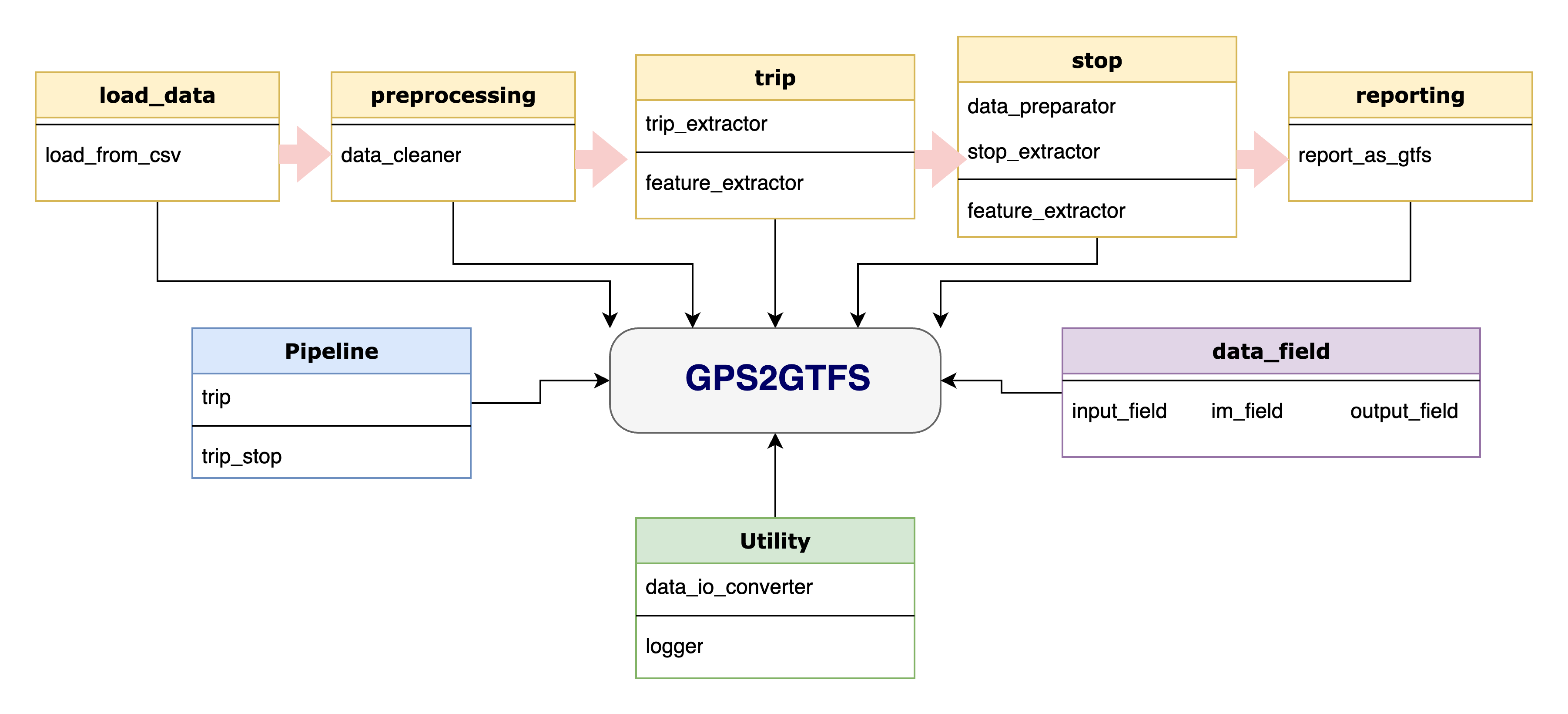}
    \caption{Software architecture diagram of the `gps2gtfs' package covering all the major functionalities of the software}
    \label{fig:soft-archi}
\end{figure}

Our `gps2gtfs' software architecture is centered around a pipeline that integrates multiple key components as shown in the figure \ref{fig:soft-archi}. It is developed in Python 3, features a modular architecture with minimal interdependence among its packages, ensuring flexibility and scalability. The package begins with the data loading module, which imports raw GPS data, followed by \textit{preprocessing}, where the data is cleaned and prepared for analysis. The \textit{trip} and \textit{stop} modules extract the key information, such as trip trajectories and matched stop sequences, and derive relevant features using specialized sub-modules. Finally, the reporting module generates the GTFS-compatible outputs. Supporting this workflow are the \textit{data\_field} module, which manages data across various stages, and \textit{utility} tools for format conversion and logging. The following sub-section explains each software function modules in a nutshell.

\subsection{Software functionalities}
\textit{  Present the major functionalities of the software.}
This sub-section presents the components of the system as that the package is structured into eight(8) primary packages:

\begin{enumerate}
    \item \textit{data\_field}: This module manages the data frame column names for user input and a predefined set of output columns ensuring the structured flow of data thorughout the software. The fields provided by the user should be a superset of the defined fields within this package.
    \item \textit{load\_data}: This module handles importing and loading the required data into the pipeline from external sources.
    \item \textit{preprocessing}: The preprocessing package is designed to clean the data loaded from the previous step using `data\_cleaner' function. Further, it removes erroneous records, ensures the continuity in the data stream and finally converts the fields in the data frame into the supported format for the main modules.
    \item \textit{trip}: One of the primary module that focus on extracting trips (trip trajectory data-points) for every bus among the pool of mixed records. This package also generating associated features for every trip which can be utilized in further analysis.
    \item \textit{stop}: The `stop' package assigned for transit stop sequence matching as described in the methodology sub-section. This outputs the the sequnces of time-stamps corresponding to arrival and departure of every stop for every trip.
    \item \textit{reporting}: This is the core module, that transforms the output from the previous modules into the recognized GTFS data format (both static and real-time) tailored for any type of transit.
    \item \textit{pipeline}: This package has the functionality to execute the trip extraction module and the stop extraction module synchronized.
    \item \textit{utility}: The utility module provides support for various utility functions, including input/output operations, conversions, and logging.
    
\end{enumerate}

\section{Illustrative examples}

This section presents some key illustrations of usage of our package and visualization in a real-world dataset. First, the python code configuration for implementing our package and the core modules such as: trip extraction (top) and the stop sequence matching (bottom) are shown below.

\lstset{
    language=Python,
    basicstyle=\ttfamily\footnotesize, 
    numbers=none,                     
    showspaces=false,
    showstringspaces=false,
    showtabs=false,
    frame=single,                     
    tabsize=4,
    captionpos=b,
    breaklines=true,
    breakatwhitespace=true,
    xleftmargin=5pt,                  
    xrightmargin=5pt,
    aboveskip=1mm,                    
    belowskip=1mm,                    
    lineskip=-1.5pt                   
}

\begin{lstlisting}[language=Python]
from gps2gtfs.pipeline.trip import run

if __name__ == "__main__":
    raw_gps_data_path = "path/to/raw_gps_data/csv"
    trip_terminals_data_path = "path/to/trip_terminals_data/csv"
    terminals_buffer_radius = 100

    run(
        raw_gps_data_path,
        trip_terminals_data_path,
        terminals_buffer_radius,
    )
\end{lstlisting}

\begin{lstlisting}[language=Python]
from gps2gtfs.pipeline.trip_stop import run

if __name__ == "__main__":
    raw_gps_data_path = "path/to/raw_gps_data/csv"
    trip_terminals_data_path = "path/to/trip_terminals_data/csv"
    stops_data_path = "path/to/stops_data/csv"
    terminals_buffer_radius = 100
    stops_buffer_radius = 50
    stops_extended_buffer_radius = 100

    run(
        raw_gps_data_path,
        trip_terminals_data_path,
        stops_data_path,
        terminals_buffer_radius,
        stops_buffer_radius,
        stops_extended_buffer_radius,
    )
\end{lstlisting}

\begin{figure}[htbp]
    \centering
    \includegraphics[width = 0.9\textwidth]{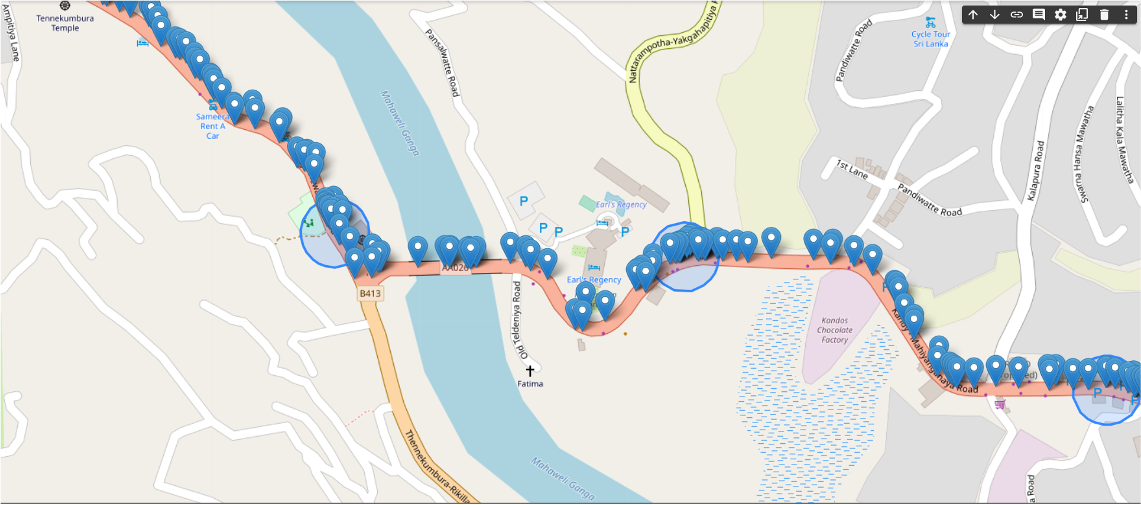}
    \caption{An illustration visualized by the 'visualization - folium' module from  the software package 'gps2gtfs' showing the GPS trajectory points collected for a single trip over its route and the buffer around bus stops to filter out corresponding data point for bus arrival and departure }
    \label{fig:stopbuffer}
\end{figure}

To illustrate the package functionalities, we used a real-world bus GPS data obtained from one of the main bus routes in Kandy City, Sri Lanka. The GPS sensors and devices were fixed in every buses in the city and the stream of GPS data collected through server and processed with our software. We developed a visualization code based upon \textit{folium} package, and inspected how the stream of GPS data are processed, trip trajectories are extracted, buffers are created on 2-D map extracted from \textit{OpenStreetMap}. One portion of the screenshot from our visualization module is shown in figure \ref{fig:stopbuffer}. The blue pins are the  extracted and matched records of a single trip from the GPS stream of data over the route (road on the 2-D map). The blue circles are the buffer created by centered around bus stops along the route. Here, multiple and sufficient records are filetered with in each bus stop buffer and hence, the arrival, departure and dwell times are extracted.

This provides an intuition on how to utilize our software package for intended downstream applications. In paricular,transit monitoring dashboard for authorities in real-time, user application to passengers for tracking and informing arrival times of next buses are some key intuitions. More impacts from our software package is explained in the following section.

\section{Impact}

This `gps2gtfs' package can spin off the advancement of reliable public transportation systems and further induce the development of transit utility software applications. This plug-and-play package contributes to the seamless integration of public transportation data into a wide range of software platforms, ultimately enhancing the overall user experience and accessibility of transit information. By simplifying the process of converting GPS data into GTFS format, this package empowers developers to create innovative solutions that improve the efficiency and effectiveness of public transportation networks. Multiple research studies have recently been conducted using GTFS static and real-time data. In Seattle, the authors of \cite{aemmer2022measurement} designed a framework that utilizes GTFS-RT data to analyze consistent and predictable delays (systematic delays) and random fluctuations on a section-by-section basis (stochastic delays). This framework serves as a tool to identify areas where reactive treatments (such as schedule padding) or proactive infrastructural modifications can be implemented. Next, Devunuri \cite{devunuri2024gtfs} developed a library called ''GTFS Segments'', which is a Python toolkit that computes, visualizes, and analyzes bus stop spacings by reading GTFS data, snapping stops to points along routes, and dividing routes into segments. In addition, visualization tools like G2Viz \cite{para2024g2viz} and PubtraVis \cite{prommaharaj2020visualizing} with six visualization modules that reflect on different transit system operational characteristics; mobility, speed, flow, density, headway, and analysis which explicitly demonstrate the potential of GTFS data. Similarly, there is significant potential to further explore the GTFS data in the coming years. Therefore, our package simplifies the conversion of raw data into the necessary GTFS format, intending to create more applications.

Moreover, this package offers more customizable parameters appropriate to the various data quality parameters (e.g., signal frequency, bus stop width) that can be adjusted based on specific needs and requirements. In the future, more modules can be added to this package, such as visualization tools to analyze transit performance, schedule adherence, delay modeling, excess dwell times, route optimization, and more. In addition, with advanced machine learning models, engineers can further develop real-time prediction models for bus arrival times, bus delays, and multi-modal trip travel time forecasting. As cities continue to invest in intelligent transportation solutions, integrating this plug-and-play package will play a key role in shaping the future of urban mobility.

The source code for our package is available on GitHub as an open-source project, which can be found at \href{https://github.com/aaivu/gps2gtfs}{https://github.com/aaivu/gps2gtfs}. Additionally, the package has been published on pypi.org and may be accessed at \href{https://pypi.org/project/gps2gtfs/}{https://pypi.org/project/gps2gtfs/}. The significant number of stars and forks indicates how widely researchers use our software. In particular, our research group is now utilizing processed GTFS data for several studies, including (1) assessing performance evaluation metrics for public transportation systems in developing countries \cite{ratneswaran2023extracting}, (2) predicting trip times of buses in diverse traffic conditions \cite{ratneswaran2023improved}, (3) public transit driver behavior analysis \cite{10689075} and other ongoing research such as Spatio-temporal graph modeling of bus travel times and prediction, integrating speed variations data, topographical and points of interest (POIs) features for bus arrival time prediction which has not yet been published.

\section{Conclusions}
To the best of our knowledge, our `gps2gtfs' package is the first available tool to practitioners to perform the transformation of GPS data to GTFS format. There has been a need for a well-designed software package that addresses all the potential challenges and with the adoptable parameters for transforming raw GPS data into GTFS format. This open-sourced 'gps2gtfs' package is designed to transform transit GPS data into GTFS real-time format under challenging circumstances like low sampling rate and non-uniformities in the data. The GTFS data format is widely used for transit data sharing and is applicable for various public transit forms. The Python package provides practical techniques for extracting crucial trip information from the raw GPS data, including sequences of trip trajectory, transit stop details, arrival and departure times at stops, dwell time at stops, trip travel duration, and running times between stops. The complex nature of mapping raw GPS data to the GTFS format is a significant hurdle especially dwell time information. The `gps2gtfs' framework is a Python 3 package designed to integrate public transportation data into various software platforms. It includes packages for data fields, load data, preprocessing, trip, stop, reporting, pipeline, and utility. The package simplifies the conversion of GPS data into GTFS format, enabling developers and researchers to create innovative solutions for improving the efficiency and effectiveness of public transportation networks. The extensible nature of the `gps2gtfs' library allows for plug-and-play interfaces that can be tailored to specific needs such as public transit-related studies, developing software applications for transit companies, authorities for real-time monitoring, advanced traveler information systems, and mainly the integration to Google Maps. The package offers customizable parameters for data quality parameters and can be expanded to include visualization tools for transit performance, schedule adherence, delay modeling, and route optimization. The package is open-sourced and can be widely used by researchers for various public transit-related studies.

\section*{Acknowledgements}
This research work was supported by the Accelerating Higher Education funded by the World Bank. [grant number - Credit/Grant \#:6026-LK/8743-LK]. We would like to thank Kajanan Selvanesan, Kesavi Aravinthan, and Gopinath Shanmugavadivel from the Department of Computer Science and Engineering, University of Moratuwa for supporting us in formatting codes and modularizing.

\bibliographystyle{elsarticle-num} 
\bibliography{reference.bib}

\end{document}